# Autonomic Resource Management in Virtual Networks

Rashid Mijumbi, Juan-Luis Gorricho and Joan Serrat
Universitat Politècnica de Catalunya
08034 Barcelona, Spain

*Abstract* — **Virtualization enables the building of multiple virtual networks over a shared substrate. One of the challenges to virtualisation is efficient resource allocation. This problem has been found to be NP hard. Therefore, most approaches to it have not only proposed static solutions, but have also made many assumptions to simplify it. In this paper, we propose a distributed, autonomic and artificial intelligence based solution to resource allocation. Our aim is to obtain self-configuring, self-optimizing, self-healing and context aware virtual networks.**

Keywords ─ Virtual Networks, Resource Allocation, Autonomic Systems, Artificial Intelligence, Context Awareness.

## I. INTRODUCTION

The Internet has advanced to such levels that its users are no longer only concerned with being able to communicate, but also by the quality and cost of this service. Achieving such user demands requires a specialization of resources and protocol stacks on the side of service providers. However, it would be economically unfeasible for each service provider to deploy independent infrastructures for every service or business context. Virtualization allows for a given network infrastructure to be shared by many service providers.

Resource Allocation is one of the main concerns in virtualisation [1]. Most approaches to this problem assume that all virtual network requirements are known in advance, and do not consider changes in network conditions, which makes them inefficient [2]. While some solutions such as [2] [3] [4] consider dynamic changes in the network, these approaches do not consider the node assignment problem [1].

We propose a distributed and autonomic [5] approach that is based on artificial intelligence [6]. Our idea is to use distributed artificial intelligence to make the virtual networks self-configuring, self-optimizing, self-healing and context aware. We propose to represent each network by an intelligent agent [7] and use Reinforcement Learning [8] to improve long term decision making capabilities and hence efficiency.

The rest of this paper is organised as follows: Sections II, III and IV present the main concepts related to the subject of our proposal, as well as a justification of our choice for this kind of solution. We present the related works in section V. Section VI explains the solution proposed in this paper, as well the scenarios that can develop from our proposed solution. We present future work and conclusions in sections VII and VIII respectively.

## II. NETWORK VIRTUALIZATION

Network virtualization is currently a subject of many research teams and has been proposed as a new technique to allow for the de-ossification of the current Internet and also facilitate new service deployment [9]. In a network virtualization environment, multiple service providers are able to create heterogeneous virtual networks (VNs) to offer customized end-to-end services to the end-users by leasing shared resources from one or more infrastructure providers (InPs) without significant investment in physical infrastructure [10]. Fig. 1, shows a representation of two virtual networks sharing resources from a single substrate network.

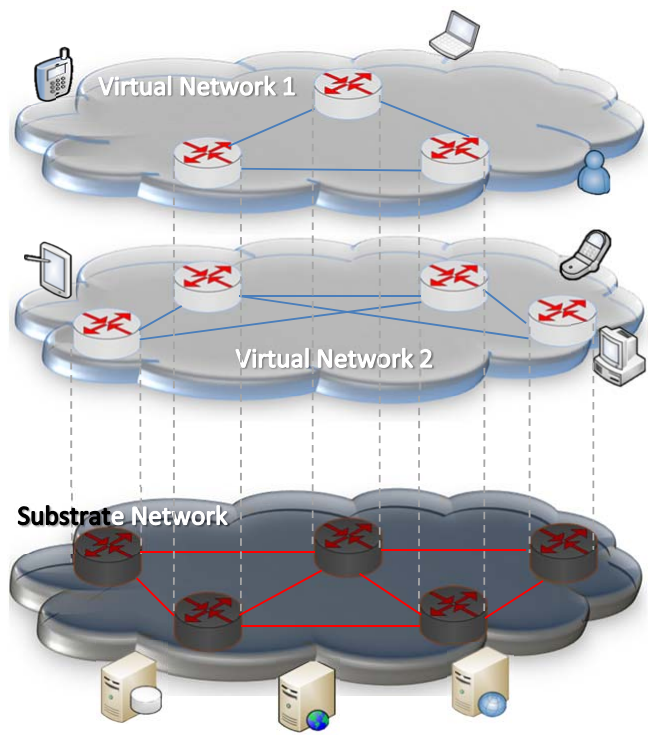

Fig. 1: Network Virtualisation

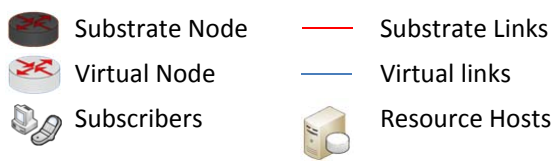

## III. AUTONOMIC SYSTEMS

Due to the growing complexity of systems, the concept of Autonomous systems [5] has recently received growing interest from both industry and Academia. The idea is to always let a given system take care of its requirements with minimal human intervention. As defined by [5], the term "autonomic" comes from an analogy to the autonomic central nervous system in the human body, which adjusts to many situations automatically without any external help. According to [5], an autonomous system should posses the different self-CHOP characteristics, which are: self-Configuring – ability to adapt to changes in the system, self-Healing – ability to recover from detected errors, self-Optimizing – ability to improve use of resources and self-Protecting – ability to anticipate and cure intrusions. We show these characteristics diagrammatically in Fig. 2.

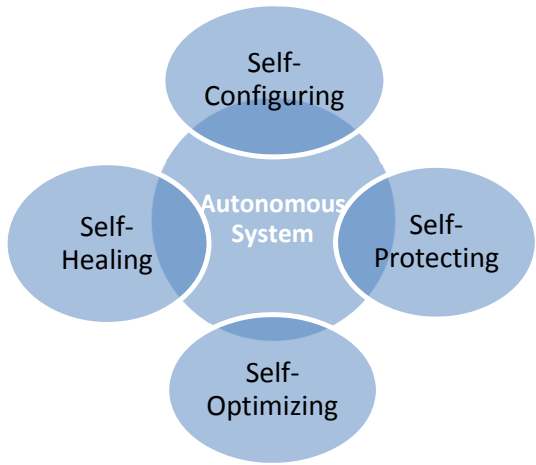

Fig. 2: Attributes of an Autonomous System

For a system to be autonomous, the actors involved should be able to manage themselves, while at the same time communicating in order to manage each other. An important aspect of such systems is that while each of them can take its own decisions, they should cooperate and coordinate at a higher level so as to achieve overall system goals and/or policies.

## IV. ARTIFICIAL INTELLIGENCE

Artificial Intelligence (AI) [6] has been defined differently by different authors. For purposes of this paper, we define AI according to [11], as "the branch of computer science that is concerned with the automation of intelligent behaviour".

In order to tackle the resource allocation problem, an interesting approach comes from Markov process [12] modeling commonly used in AI. In fact, we could achieve more accurate modeling by using Markov Decision Processes (MDPs) in which case we influence how the system makes decisions in transiting from a given state to another. In this case, the aim is always to establish a policy for system transition from one state to another. One alternative for learning a policy, also considered one of the main research branches within AI, is the reinforcement learning technique.

### A. Reinforcement Learning

Reinforcement Learning (RL) is a part of Artificial Intelligence in which an agent is situated in an environment and is entrusted with taking autonomous decisions so as to maximize the reward it gets from the environment [8]. RL is a feedback based approach where an agent receives an immediate reward for its previous action, and from there on, it will try to learn a better policy for the long run, in order to maximize a given utility function. Therefore, the agent's goal – roughly speaking – is to maximize the total amount of reward it receives in the long run [8].

Fig. 3 shows an interaction between an agent and an environment in a typical Reinforcement Learning situation. The Agent while in a given state $s_t$ performs action $a_t$ and gets a reward $r_t$ from the environment. Reinforcement learning methods specify how the agent changes its policy as a result of its experience.

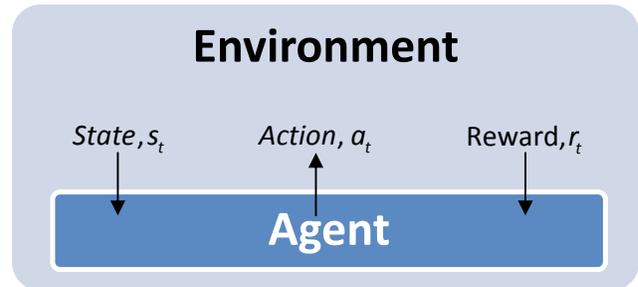

Fig. 3: Agent – Environment interactions in Reinforcement Learning

Reinforcement Learning has been proposed [13] for a resource allocation problem with capacity constraints, while [14] has used a Reinforcement Learning approach to solve a problem that is characterised as NP hard. In each case, improvements have been reported as compared to previous solutions.

### B. Multi-Agent Systems

A multi-agent system (MAS) can be defined as a group of autonomous, interacting entities sharing a common environment, which they perceive with sensors and upon which they act with actuators [15]. Depending on the application, the interaction between the agents in a MAS can either be cooperative or competitive.

MASs have – for some time now – been a subject of research for applications in fields which require autonomous and intelligent behaviour such as Robotics, distributed control and telecommunications [7]. Multi-agent systems are ideal for problems that require autonomous decision making capabilities [16].

## V. RELATED WORK

The resource allocation problem is NP hard [10]. Therefore, it has been addressed mainly using heuristic solutions. While heuristics and meta-heuristics are useful for very large instances of the problem, it should be noted that there are still challenging issues in distribution, survivability and cooperation that must be considered in these kinds of problems [9].

[10], [17] and [18] present static solutions to resource allocation, in a way that they assume that all the requests of the different virtual networks are known in advance.

[19] proposes path splitting and migration to improve virtual network resource allocation. However, the authors assume that resource requests can be delayed for some time. [20] presents a solution which is "partially" dynamic as it only allows for reconfiguration of the most critical virtual networks or services. It also assumes infinite capacity of substrate nodes and links so as to avoid the effects of rejected requests from virtual networks. [21] proposes a re-optimization solution that can be considered dynamic. However, the solution is reactive, in a way that re-optimization is only carried out after rejections.

[22] proposes a distributed resource management approach but restricts the network topology. [23] proposes a framework and algorithm that aims to reduce resource contention by reducing the loading of the links. However, the solution is also based on a specific network topology, does not consider changes in customer and network conditions and the allocation mechanism is based on central server decisions. [24] also considers a specific virtual network topology.

Artificial intelligence techniques have been proposed by [25] for the design of an efficient, dynamic and autonomic network management system. In particular, multi-agent systems are rapidly finding applications in a variety of domains, including robotics, distributed control, telecommunications, and economics [15]. [26] proposes a multi-agent based approach to autonomic resource management, while [13] proposes reinforcement learning for dynamic resource allocation in a telecommunication network.

The solutions above make simplifying assumptions such as unbounded network resources or a fixed topology, and most are static. None of the solutions provides an autonomic resource allocation solution which is self-configuring, self-optimizing, self-healing and context aware.

## VI. PROPOSED SOLUTION

To face the complexity of the problem we are faced with, we propose to decentralize the solution such that the virtual as well as substrate networks are represented by intelligent agents. Specifically, we propose that the agents use RL for resources allocation.

### A. Scenario-based Approach

Our approach to the resource allocation problem is to develop a solution that is initially based on specific scenarios, and later on extended to general contexts. In Fig. 4, we present a problem case.

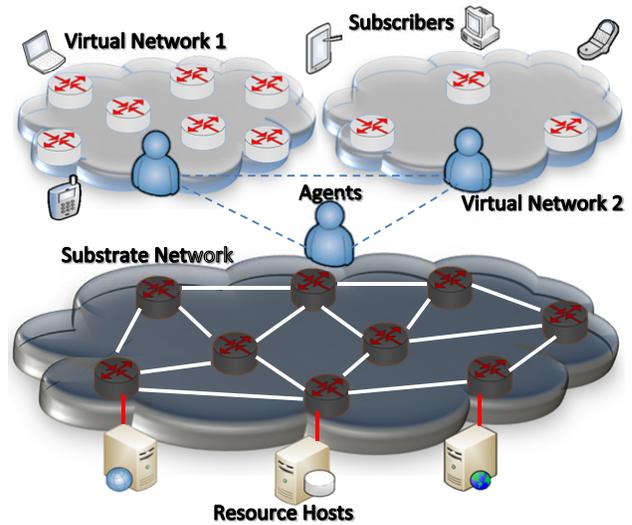

Fig. 4: Virtual Network Environment Proposed Framework

In our solution scenario, the physical nodes and links have limited capacities such as CPU capacity and link bandwidth, and the all the agents know these constraints.

Each virtual network agent (VNA) always has information on the loadings of the substrate network, which is supplied by the substrate network agent (SNA). The objective of SNA is to minimise the number of virtual network requests rejected, and ensure that based on traffic, the links and nodes have balanced loads. On the other hand, the objective of each VNA is to minimise the costs they incur (resulting from use of SN resources), while at the same time ensuring satisfaction of their users. The agents can communicate amongst themselves, and can carry out negotiations.

Users send requests for service to the virtual networks. The users, in their requests specify the nature of service required (which will in effect determine the final destination) as well as bandwidth requirements.

With the above features in consideration, we now present some of the cases that can result from our model:

*I) Self-Configuration:* As users make requests for resources, VNAs continuously evaluate their network topologies to establish if they could do better considering the status of the substrate network. Whenever they find possibilities for optimization, they can make requests for this to the SNA. VNAs can also negotiate and cooperate amongst themselves so as to agree on the usage of substrate resources and achieve the best utility both on individual and system levels

*II) Self-Optimization:* As the conditions of the substrate network change, SNA should exhibit proactive as well as reactive characteristics. For example, if a physical node and/or link is added to the network, or if their capacities are changed, SNA should re-evaluate the network loadings to establish possibilities for re-allocation of resources. While this would ideally not require any action on the part of the VNAs, in our solution we require that these agents always look out for possibilities of optimizing their resource usage whenever there are changes to the substrate network.

*III) Self-Healing:* If for any reason a node and/or link becomes un available, SNA – in collaboration with VNAs – should make decisions so as to cause the minimal possible disruptions in customer services. This situation is different from II) in a way that while self-optimization is mainly aimed at optimizing costs, and possible improvements in customer service levels, self-healing is much more urgent as in such cases there are possibilities of violating agreements with customers. Therefore, if the VNAs have to cooperate and/or negotiate in one of these two cases, an agent will adjust its objectives considering the possible conflict between utility optimisation and customer satisfaction.

*IV) Context Awareness:* Whenever users are connected to the network, they will send periodic updates about their location. Based on these locations, VNAs will determine not only the exact location of the customer, but may also reason about the actual state of the user, for example; the agent can know if a user is walking, driving or stationary. Based on this context information, the virtual network may take decisions about the resources being used by the user. For example, a user whose location is near a Wi-Fi hot spot **AND** this user is stationary **AND** this user is occupying a high bandwidth – say for a video on demand service – could be offloaded to the Wi-Fi and if there are necessary changes to the nodes for this specific customer, these changes can be effected by SNA.

## VII. FUTURE WORK

This paper presents a work in progress. Over the course of this work, we will propose a resource allocation algorithm based on reinforcement learning, as well as a negotiation protocol based on network and user context to allow for the coordinated and profitable use of substrate network resources.

We will also propose a RL based utility model on the basis of which intelligent agents that represent the different network owners make autonomous decisions aimed at efficient and optimal resource allocation. To allow for reasoning and hence intelligent decision making, a representation of the different instances of our model will also be proposed. To this end, if we deem it necessary, an ontology representing these instances will be proposed. Lastly, we will carry out simulations to prove our concept.

## VIII. CONCLUSION

In this paper, we have proposed an autonomic and distributed solution to virtual network resource management. We have proposed that each virtual network can be represented by an intelligent each with individual objectives, and that these agents can cooperate with each other so as to achieve overall system objectives. In order to solve the space and time complexities that are associated with this problem, we have proposed a reinforcement learning approach such that the agents learn to perform better over time.

## ACKNOWLEDGMENT

This work has been started within the project TEC2009-14598-C02-02 granted by the MEC Spanish Ministry and partially funded with FEDER funding. We also want to acknowledge the EVANS EU project funded under contract PIRSES-GA-2010-269323.